\RequirePackage[svgnames]{xcolor}

\documentclass[11pt,letterpaper]{mystyle}

\usepackage[all]{hypcap}
\usepackage[svgnames]{xcolor}
\usepackage[comma,authoryear,compress]{natbib}
\usepackage{hyperref}[citecolor=lightblue]

\hypersetup{
    colorlinks = true,
    citecolor = {YaleBlue},
}

\usepackage{algorithm}
\usepackage{algorithmicx}
\usepackage{algpseudocode}
\usepackage{microtype}
\expandafter\def\csname ver@subfig.sty\endcsname{}
\usepackage{booktabs} %
\usepackage{float}
\usepackage{bigstrut}

\usepackage{amsmath}
\usepackage{amssymb}
\usepackage{mathtools}
\usepackage{amsthm}
\usepackage{mathrsfs}
\usepackage{nicefrac}
\usepackage{dsfont}
\usepackage{enumitem}
\usepackage{cleveref}
\usepackage{bxcoloremoji}
\usepackage{float}
\usepackage[utf8]{inputenc} %
\usepackage[T1]{fontenc}    %
\usepackage{hyperref}       %
\usepackage{url}            %
\usepackage{booktabs}       %
\usepackage{amsfonts}       %
\usepackage{nicefrac}       %
\usepackage{microtype}      %
\usepackage{graphicx}
\usepackage{subcaption} 
\usepackage{subfig}
\usepackage{amssymb}
\usepackage{fdsymbol}
\usepackage{wrapfig}
\usepackage{lipsum}
\usepackage{stackengine}
\usepackage[font=small,labelfont=bf]{caption}
\usepackage{color}
\usepackage{adjustbox}

\usepackage{rotating}
\usepackage{makecell}
\usepackage[absolute,overlay]{textpos}

\newcommand{\x}{\mathbf{x}}

\definecolor{blanchedalmond}{rgb}{1.0, 0.92, 0.8}
\definecolor{carmine}{rgb}{0.59, 0.0, 0.09}
\definecolor{lightblue}{rgb}{0.22,0.45,0.70}%

\renewcommand{\mathbf}{\boldsymbol}

\makeatletter
\def\Ddots{\mathinner{\mkern1mu\raise\p@
\vbox{\kern7\p@\hbox{.}}\mkern2mu
\raise4\p@\hbox{.}\mkern2mu\raise7\p@\hbox{.}\mkern1mu}}
\makeatother

\definecolor{amaranth}{rgb}{0.9, 0.17, 0.31}
\definecolor{antiquebrass}{rgb}{0.8, 0.58, 0.46}
\definecolor{antiquefuchsia}{rgb}{0.57, 0.36, 0.51}
\definecolor{chromeyellow}{rgb}{0.31, 0.47, 0.26}

\newcommand{\github}{\raisebox{-1.5pt}{\includegraphics[height=1.05em]{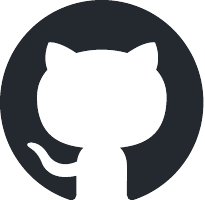}}}

\newtcolorbox{AIbox}[2][]{aibox,title=#2,#1}
\definecolor{lightblue}{rgb}{0.22,0.45,0.70}%
\definecolor{Gray}{gray}{0.95}
\definecolor{Cornsilk}{rgb}{1.0, 0.97, 0.86}

\setlist[itemize]{leftmargin=*}
\tcbuselibrary{listingsutf8}

\definecolor{jsonbg}{RGB}{245, 245, 250}       
\definecolor{jsonframe}{RGB}{100, 100, 120}    
\definecolor{jsontitle}{RGB}{0, 0, 0}          
\definecolor{jsonkey}{RGB}{130, 60, 60}     
\definecolor{jsonvalue}{RGB}{0, 0, 150}        
\definecolor{jsoncomment}{RGB}{77, 77, 77}     

\lstdefinelanguage{json}{
    basicstyle=\ttfamily\small,
    showstringspaces=false,
    breaklines=true,
    frame=none,
    backgroundcolor=\color{jsonbg},
    string=[b]",
    morestring=[b]',
    stringstyle=\color{jsonkey},
    commentstyle=\color{jsoncomment},
    keywordstyle=\color{jsonvalue},
    morekeywords={true,false,null}
}

\newtcblisting{jsonlisting}[1][]{
  enhanced,
  listing only,
  listing options={   
    language=json,
    basicstyle=\ttfamily\small,
    breaklines=true,
    showstringspaces=false,
    frame=none,
    backgroundcolor=\color{jsonbg},
    keywordstyle=\color{jsonvalue},
    stringstyle=\color{jsonkey},
    commentstyle=\color{jsoncomment}
  },
  colback=jsonbg,
  colframe=jsonframe,
  coltitle=jsontitle,
  fonttitle=\bfseries,
  top=5pt,
  bottom=5pt,
  left=5pt,
  right=5pt,
  sharp corners,
  boxrule=0.5pt,
  title style={top color=white, bottom color=jsonbg},
  #1
}

\title{Planet as a Brain: Towards Internet of AgentSites based on AIOS Server}

\runningtitle{Planet as a Brain: Towards Internet of AgentSites based on AIOS Server}

\author{
  Xiang Zhang$^{\dagger\ast}$,
  Yongfeng Zhang$^{\ddagger\ast}$\\
  $^{\dagger}$Boston University, $^{\ddagger}$Rutgers University, $^{\ast}$AIOS Foundation\\
  \href{mailto:research@aios.foundation}{research@aios.foundation}
}

\correspondingauthor{Xiang Zhang, Department of Computer Science, Metropolitan College, Boston University, Boston, MA 02215, \href{mailto:xz0224@bu.edu}{xz0224@bu.edu}; Yongfeng Zhang, Department of Computer Science, Rutgers University, New Brunswick, NJ 08901, \href{mailto:yongfeng.zhang@rutgers.edu}{yongfeng.zhang@rutgers.edu}}

\begin{document}


\setlength{\TPHorizModule}{1cm}
\setlength{\TPVertModule}{1cm}
\begin{textblock}{5}(0,0) 
    \includegraphics[width=1cm]{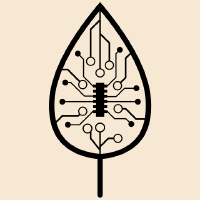}
    \raisebox{0.618\height}{{{\Large AIOS Foundation}}}
\end{textblock}

\begin{abstract}
The internet is undergoing a historical transformation from the ``Internet of Websites'' to the ``Internet of AgentSites.'' While traditional Websites served as the foundation for information hosting and dissemination, a new frontier is emerging where AgentSites serve as the hubs of the internet, where each AgentSite hosts one or more AI agents that receive tasks, address them, and deliver actionable solutions, marking a significant shift in the digital landscape and representing the next generation of online ecosystems. Under this vision, AIOS, the AI Agent Operating System, serves as the server for the development, deployment and execution of AI agents, which is a fundamental infrastructure for the Internet of Agentsites.\\

In this paper, we introduce AIOS Server, a runtime framework designed to host agents and support large-scale collaboration among decentralized agents. AIOS Server provides a standardized communication protocol leveraging the Model Context Protocol (MCP) and JSON-RPC to enable structured agent-agent or human-agent interactions. Each AIOS node operates as an independent server, capable of hosting and executing agents, while supporting peer-to-peer coordination without reliance on centralized orchestration. \\

Based on the AIOS Server, we further present the world's first practically deployed Internet of Agentsites (AIOS-IoA)\textsuperscript{1}, including AgentHub for agent registration and management as well as AgentChat for interactive communication. Furthermore, we prototype an agent discovery mechanism based on Distributed Hash Tables (DHT) and a Gossip protocol, which serves as the search engine for the internet of agentsites, enabling scalable and resilient agent registry and lookup on this new internet.\\

Our evaluation demonstrates that AIOS Server achieves low-latency communication, efficient task delegation, and robust coordination in agent networks. This work provides a practical foundation for building the Internet of Agentsites — a new paradigm where autonomous agents become first-class citizens of the web. The implementation is available at GitHub\textsuperscript{2}.
\vspace{5mm}


\coloremojicode{1F3E0} \textbf{$^1$Website}: \href{https://planet.aios.foundation}{https://planet.aios.foundation}

\github{} \textbf{$^2$Code Repository}: \href{https://github.com/agiresearch/AIOS}{https://github.com/agiresearch/AIOS}, \href{https://github.com/agiresearch/AIOS.Server}{https://github.com/agiresearch/AIOS.Server}





\end{abstract}

\maketitle
\vspace{3mm}

\section{Introduction}

The rapid progress of large language models (LLMs) has led to the emergence of autonomous agents capable of planning, reasoning, and interacting with humans and other agents in structured environments \citep{wang2024a, yao2022, xu2025a, wang2024b}. These agents increasingly act beyond static response generation, demonstrating memory, tool usage, and long-term task execution abilities. However, most existing agent-based systems remain confined within centralized platforms, limiting openness, interoperability, and scalability in multi-agent ecosystems.

To address these limitations, this paper introduces AIOS Server — a runtime framework designed to host autonomous agents and facilitate large-scale agent communication across the internet. Each AIOS server hosts one or more autonomous AI agents, constituting an Agentsite — just like the way that Nginx server hosts Websites. Agentsites are further connected through the internet for human-agent and agent-agent communication, establishing the Internet of Agentsites, akin to the Internet of Websites that constitute the world wide web (WWW). 

AIOS Server provides a standardized communication protocol combining the MCP (Model Context Protocol) and JSON-RPC (JavaScript Object Notation - Remote Procedure Call), enabling structured interactions between agents, humans, and external services. Each AIOS Server node operates as an independent execution environment, supporting dynamic agent deployment, peer-to-peer communication, and decentralized coordination without relying on centralized control.

Building on the AIOS Server, we further present the world's first practically deployed Internet of Agentsites (AIOS-IoA) — an open ecosystem where distributed agents, hosted across heterogeneous sites (AgentSites), interact and collaborate at internet scale. The AIOS-IoA architecture includes AgentHub for agent registration and management, and AgentChat for interactive human-agent communication, both available at \url{https://planet.aios.foundation}.

To enable scalable and resilient agent discovery across this decentralized network, we design and prototype an agent search mechanism based on Distributed Hash Tables (DHT) and a Gossip protocol. This decentralized search engine provides efficient registry, lookup, and discovery functionalities, allowing agents to interact across distributed environments in a robust and fault-tolerant manner.

\begin{wrapfigure}{r}{0.45\textwidth}
\vspace{-2ex}
  \centering
  \includegraphics[width=0.45\textwidth]{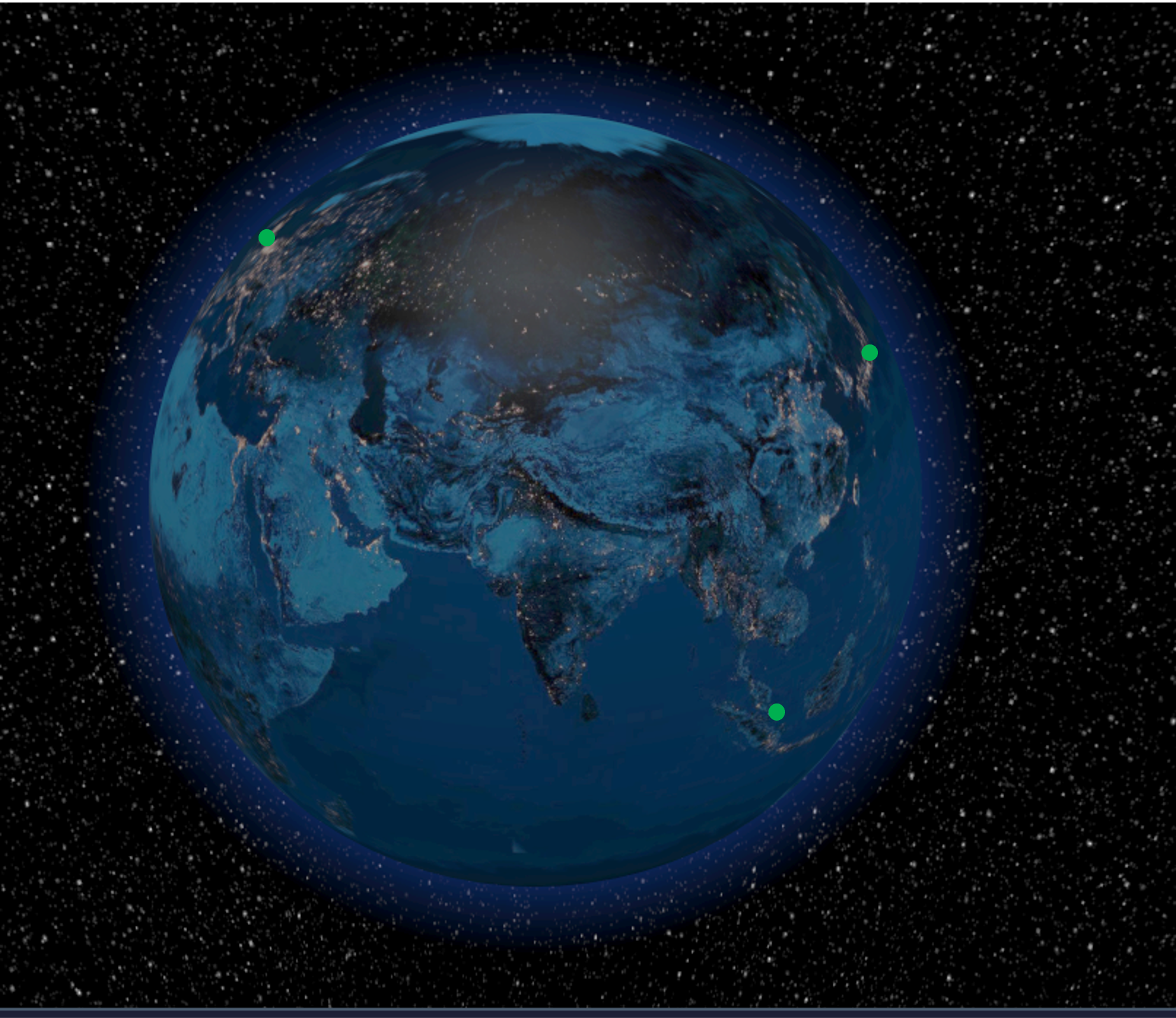}
  \caption{Global view showing agentsites deployed in London, Singapore, and Tokyo.}
  \label{fig:wrapfig}
  \vspace{-8ex}
\end{wrapfigure}

In summary, our key contributions are: 
\begin{itemize} 
\item We propose AIOS Server, a decentralized runtime framework enabling structured communication and coordination among autonomous agents. 
\item We implement the Internet of Agentsites (AIOS-IoA), providing the first practical deployment of an open, agent-centric web ecosystem. 
\item We design and evaluate a DHT-based decentralized agent registration and discovery mechanism, supporting scalable search and registry across agent networks. 
\item We empirically evaluate AIOS Server in real-world deployment settings, demonstrating low-latency communication, efficient task delegation, and robust peer-to-peer coordination. 
\end{itemize}

This work builds on recent advances in agent system infrastructures \citep{mei2024aiosllmagentoperating, hong2023, deng2024}, memory-enhanced LLM agents \citep{xu2025a, wang2024b}, and agent communication protocols \citep{chan2023, du2023, liang2023}. AIOS Server emphasizes modularity, interoperability, and robustness as foundational elements for open and scalable agent ecosystems. This work contributes a practical foundation for building the next-generation Internet of Agentsites, where autonomous agents operate as first-class citizens of the web, capable of decentralized collaboration, search, and interaction. 

Recent works such as \cite{chen2024internetagentsweavingweb} explored the idea of the internet of agents as a multi-agent framework in terms of agent team up and collaboration. However, our work is different in that we do not focus on agent team up and collaboration, but on the physical, decentralized and real-world deployment of agents across the globe (as shown in Figure \ref{fig:wrapfig}), just like the physical Internet of Websites, where each website runs on a server at a certain location on the planet, and servers are connected through the internet to constitute the world wide web (WWW). As a result, our infrastructure is denoted as the Internet of Agentsites for differentiation and for highlighting the analogy with the Internet of Websites.

\section{Related Work and Background}

\subsection{LLM-Powered Agent Systems}

Recent advances in large language models (LLMs) have enabled the creation of autonomous agents that can interpret tasks, use tools, and maintain state over long interactions \citep{wang2024a, yao2022, zhou2023agents, yu2023musicagent}. These agents are no longer limited to single-turn responses—they can reason, remember, and act in structured workflows \citep{xu2025a, wang2024b}. Frameworks such as Camel \cite{li2023camel}, OpenAGI \cite{ge2023openagi}, AIOS \cite{mei2024aiosllmagentoperating}, AutoGen \citep{wu2023} and MetaGPT \citep{hong2023} define agent execution workflows and pipelines, where agents assume roles to work on task decomposition and decision-making either individually or collaboratively. Memory-augmented systems, including A-Mem and Workflow Memory \citep{xu2025a, wang2024b}, provide persistent state tracking, improving contextual understanding and coherence. LLM-based agents now actively support applications in software development \citep{qian2023}, web interaction \citep{iong2024, deng2024}, and interactive simulations \citep{wang2023a}, extending their practical utility across domains.

\subsection{Agent Communication and Protocol Design}

Effective coordination in agent systems requires clear, interpretable communication. Structured interaction protocols—such as ReAct \citep{yao2022} and intent-based dialogue models—enable agents to plan, reason, and act based on conversational or environmental context. Recent works explore debate and discussion as communication primitives to improve factuality and reasoning depth \citep{du2023, liang2023, chen2024internetagentsweavingweb, pang2024, wu2023}. Message-oriented protocols like JSON-RPC are widely adopted to support structured, machine-readable interaction between agent components or subsystems. Emerging platforms such as AgentVerse \citep{chen2023}, OS-Copilot \citep{wu2024b}, and Formal-LLM \citep{li2024} emphasize modular communication layers, enabling agents to interact with tools, APIs, and other services in well-defined formats.

\subsection{Infrastructure and Decentralized Execution}

Traditional AI agents often depend on centralized backends for registration, execution, and orchestration. This architecture limits scalability, introduces single points of failure, and reduces system adaptability. Recent efforts propose decentralized execution environments that distribute agent hosting and discovery across independent nodes. Platforms like AIOS (AI Agent Operating System) \citep{mei2024aiosllmagentoperating} highlight the need for developer-friendly, general-purpose agent development, hosting and execution frameworks. Our system builds on this foundation by introducing a runtime that integrates agent communication, registration, and discovery into a fully distributed infrastructure. Security and evaluation remain key concerns. Benchmarking frameworks such as ASB \citep{zhang2024a} and GAIA \citep{mialon2023} provide standardized settings for assessing agent robustness, safety, and coordination capabilities at scale.

\section{Overview of System Architecture}

\begin{figure}[t]
    \centering
    \includegraphics[width=1.0\linewidth]{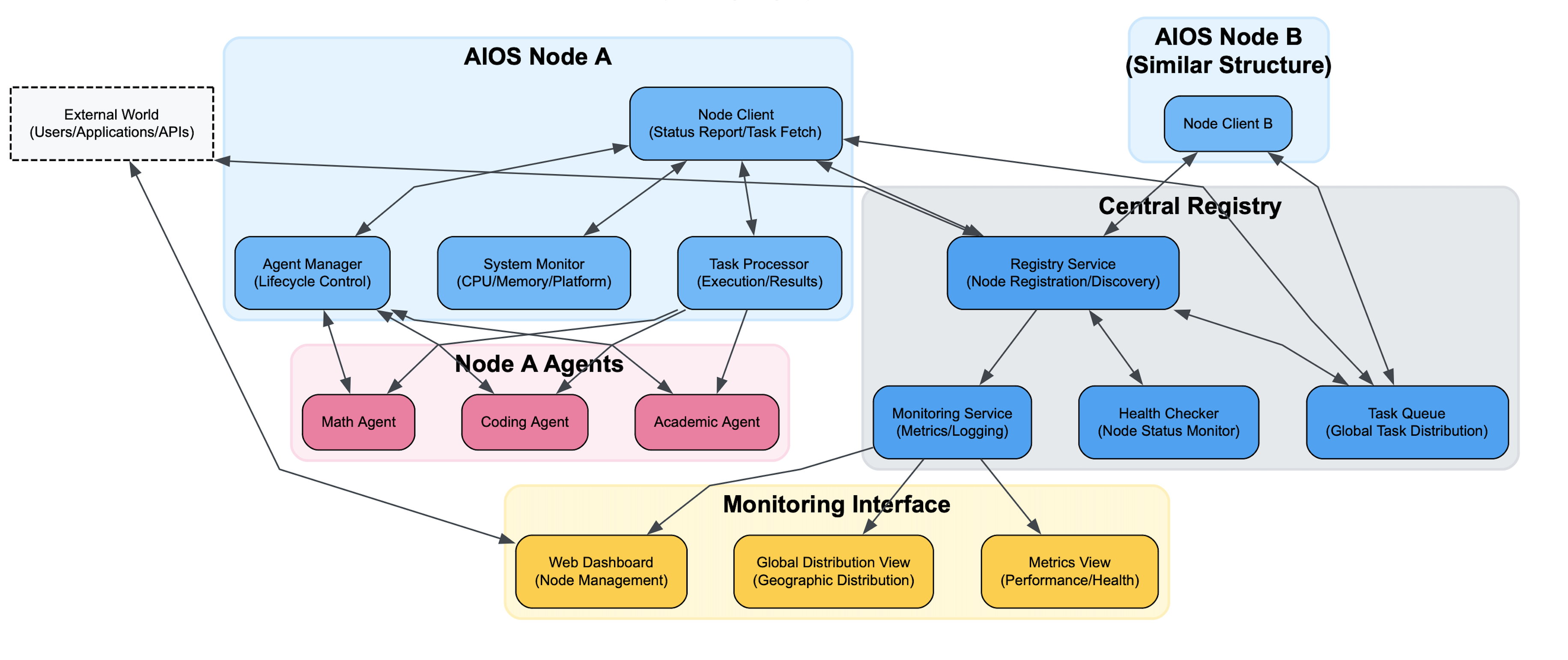}
    \vspace{-20pt}
    \caption{AIOS Server architecture with layers for messaging, agents, and services.}
    \label{fig:ioa_architecture}
    \vspace{-10pt}
\end{figure}

AIOS Server operates as a layered architecture that facilitates communication between human users and autonomous agents. 
Figure~\ref{fig:ioa_architecture} illustrates the overall architecture of the AIOS server system. 

Each AIOS server node operates independently, hosting agents and managing tasks. Core components include an agent manager, system monitor, task processor, and node client.
AIOS server nodes interact with an agent registry node for agent registration, discovery, task assignment, and health monitoring. The registry node maintains the metadata of AIOS server nodes and provides an interface for users to interact and delegate tasks to the agents on various AIOS servers. This node registration design enhances the safety and security of the Internet of Agentsites since malicious agents can be discovered at the registry node to prevent harms to the network. Note that there can be more than one registry nodes in the network and each AIOS server node can decide which registry node(s) to register itself. Furthermore, a web-based monitoring interface at the registry provides real-time visibility into node performance, geographic distribution, and system metrics.

In the following sections, we will introduce the agent communication, registration, discovery, and execution protocols step by step.

\section{AIOS Communication Protocol}

We start by designing the AIOS Communication Protocol, which facilitates structured interactions in agent-based systems, encompassing \textit{(i)} Human-Agent Communication Protocol and \textit{(ii)} Agent-Agent Communication Protocol. To ensure interoperability and scalability, we implement based on the Model Context Protocol (MCP) v1.2.1\footnote{\url{https://modelcontextprotocol.io/introduction}}, a standardized framework for integrating Large Language Models (LLMs) with external tools and data sources.

MCP follows a client-server architecture, where structured requests and responses enable seamless AI-driven workflows. LLMs often require access to external computation, structured data, and APIs to improve reasoning and task execution. Traditional integrations rely on \textit{ad hoc} solutions, limiting scalability and security. MCP addresses these limitations by providing:

\begin{itemize}
    \item \textit{Interoperability}: A standardized interface compatible with multiple LLM providers;
    \item \textit{Modular architecture}: Decoupled deployment of models, tools, and data sources;
    \item \textit{Secure data handling}: Controlled access aligned with infrastructure constraints.
\end{itemize}

We present an MCP v1.2.1 implementation for agent-agent and human-agent communication, using JSON-RPC for structured request-response exchanges. We explore its role in AI workflow orchestration, inter-agent collaboration, and scalable task delegation within the AIOS-IoA framework.

\subsection{Human-Agent Communication Protocol}

The Human-Agent Communication Protocol enables structured interaction between humans and AI agents, where human users can issue tasks, request information, and receive structured responses. The agents running on AIOS server operate as intelligent assistants that interpret user requests, perform computations, and return results in a standardized format.

The MCP-based communication workflow follows a structured protocol to ensure efficient and scalable interactions. The process consists of four key steps:
\begin{enumerate}[leftmargin=*]
    \item \textit{Task Initialization}: A structured JSON-RPC request is issued to an MCP-compliant agent.
    \item \textit{Processing}: The agent interprets the request and executes the assigned task.
    \item \textit{Response Generation}: The agent returns a structured response.
    \item \textit{Iterative Refinement (Optional)}: The requester may refine the query, triggering further interactions.
\end{enumerate}

\begin{figure}[t]
    \centering
    \includegraphics[width=0.9\linewidth]{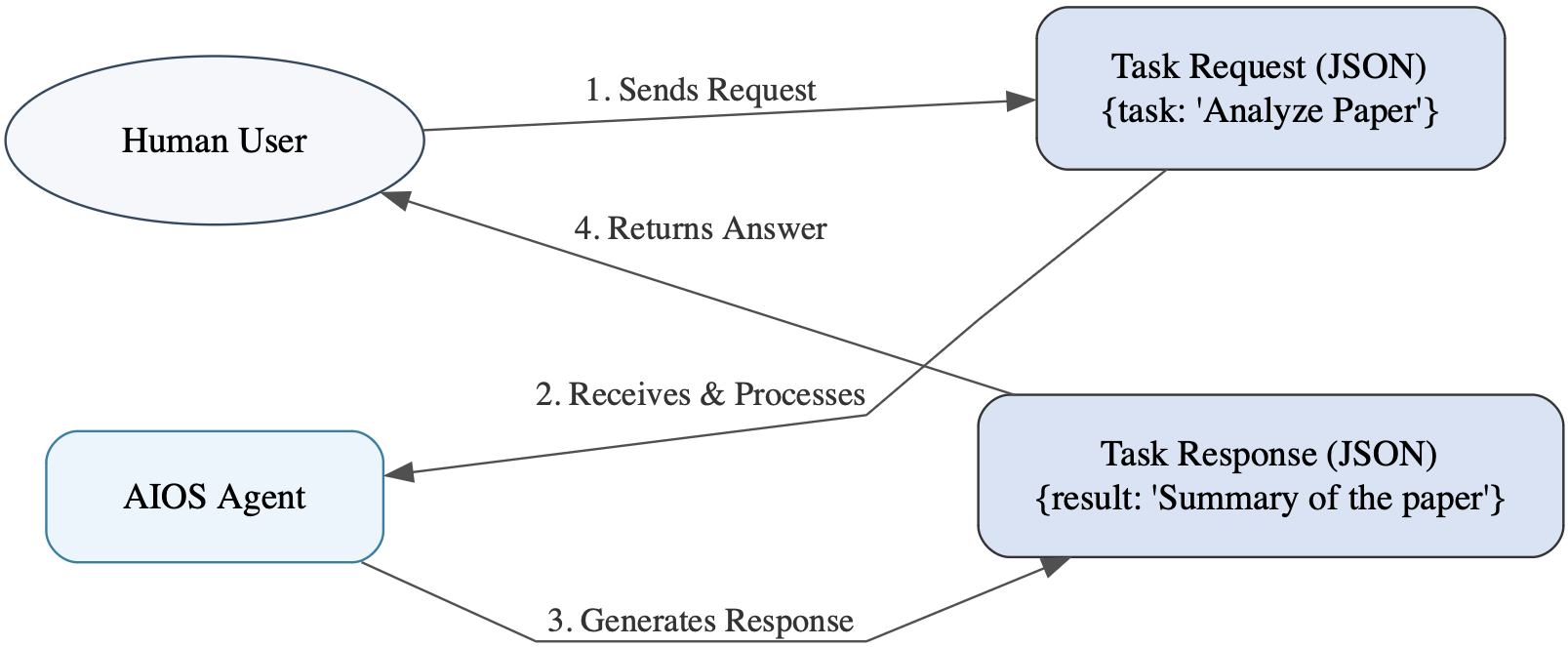}
    \caption{Human-Agent Protocol: Users communicate with AI agents using structured requests.}
    \label{fig:Human-Agent}
\end{figure}

Figure~\ref{fig:Human-Agent} shows how human users interact with agents through standardized workflow. Requests are submitted as structured prompts and agents return contextual responses via the MCP protocol. The communication workflow possesses the following key features: 1) \textit{Standardized Message Format}, which ensures structured, machine-readable exchanges; 2) \textit{Context-Aware Execution}, where agents process requests with system and conversational context awareness; 3) \textit{Multi-Turn Capability}, which supports iterative refinements and follow-up queries; 4) \textit{Progress and Error Handling}, which enables structured status tracking and exception management. For better illustration, we provide examples of user request in Appendix~\ref{appendix:human-agent-request}, and examples of AI agent response in  Appendix~\ref{appendix:human-agent-response}.

\subsection{Agent-Agent Communication Protocol}

The Agent-Agent Communication Protocol facilitates interactions between AI agents in a decentralized, distributed computing environment. Agents communicate dynamically to delegate tasks, share data, and execute workflows collaboratively. The communication workflow consists of four key steps:
\begin{enumerate}[leftmargin=*]
    \item \textit{Agent Discovery}: Agents dynamically identify peers via a distributed registry.
    \item \textit{Task Delegation}: An agent delegates a task to another capable agent.
    \item \textit{Task Execution}: The receiving agent processes the task.
    \item \textit{Response Handling}: The requesting agent integrates the results into its workflow.
\end{enumerate}

\begin{figure}[t]
    \centering
    \includegraphics[width=0.9\linewidth]{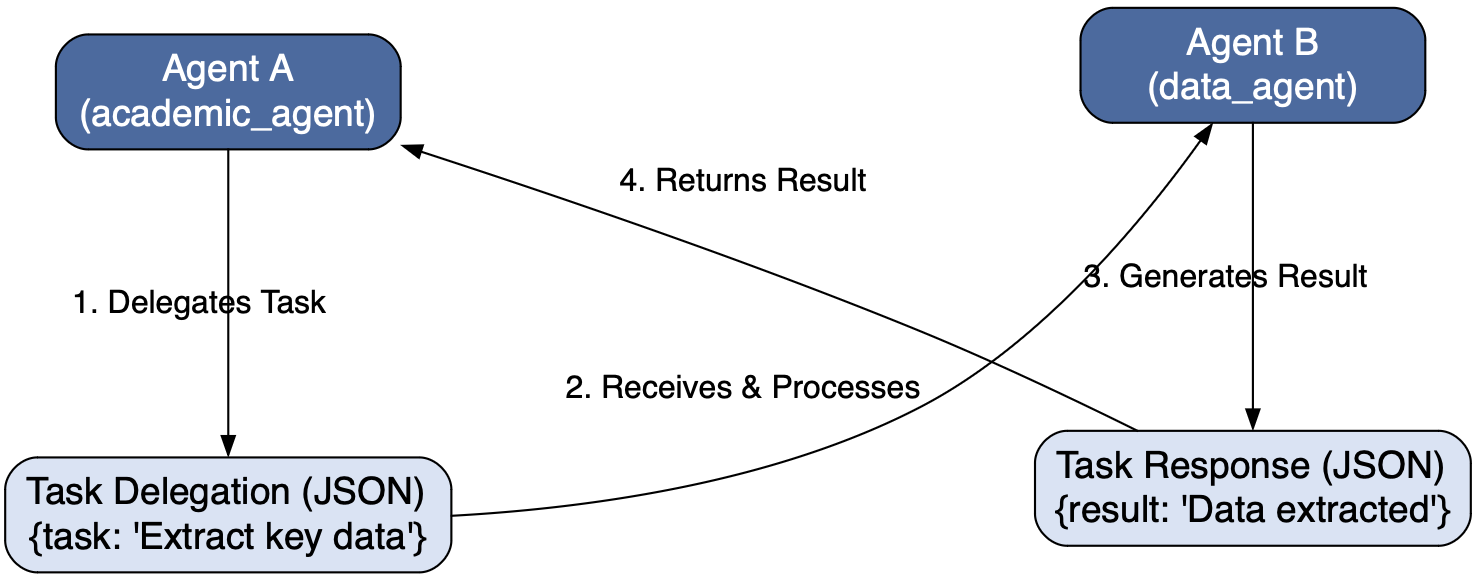}
    \caption{Agent-Agent Protocol: Structured messaging between autonomous agents.}
    \label{fig:Agent-Agent}
\end{figure}

Figure~\ref{fig:Agent-Agent} illustrates the agent-to-agent communication process. Agents exchange structured messages using JSON-RPC to delegate and coordinate tasks. The protocol enables multi-stage workflows across distributed nodes. The communication workflow possesses the following key features: 1) \textit{Decentralized Agent Lookup}, which is realized through dynamic agent discovery via distributed registries; 2) \textit{Intent-Based Messaging}, where agents can specify their roles (query, delegate, collaborate) when sending messages through the internet; 3) \textit{Hierarchical Task Execution}, which enables multi-step workflows across agents; 4) \textit{Message Routing}, which is naturally supported by IP-based message routing and delivery through the internet. Similarly, examples of task delegation request are provided in Appendix~\ref{appendix:agent-request}, and examples of agent response are provided in Appendix~\ref{appendix:agent-response}.

\subsection{Human-Agent vs. Agent-Agent Communication}

For better understanding of the similarity and difference between the human-agent and agent-agent communication protocols, we summarize the key features of the two protocols in Table \ref{tab:protocol_comparison}.
\begin{table}[t]
\centering
\begin{tabular}{|c|c|c|}
\hline
\textbf{Feature} & \textbf{Human-Agent Communication} & \textbf{Agent-Agent Communication} \\
\hline
\textit{Initiator} & Human user & AI agent \\
\hline
\textit{Message Flow} & Request $\rightarrow$ Response & Request $\rightarrow$ Task Delegation $\rightarrow$ Response \\
\hline
\textit{Interaction Type} & Direct command execution & Autonomous collaboration \\
\hline
\textit{Capabilities} & Single-task execution & Multi-task delegation \\
\hline
\textit{Routing} & Direct & Dynamic peer-to-peer \\
\hline
\textit{Use Case} & AI assistants, task execution & Distributed AI, multi-agent workflows \\
\hline
\end{tabular}
\caption{Comparison of human-agent and agent-agent communication protocols}
\label{tab:protocol_comparison}
\end{table}

\subsection{Communication Examples}

\begin{figure}[t]
    \centering
    \includegraphics[width=1.0\linewidth]{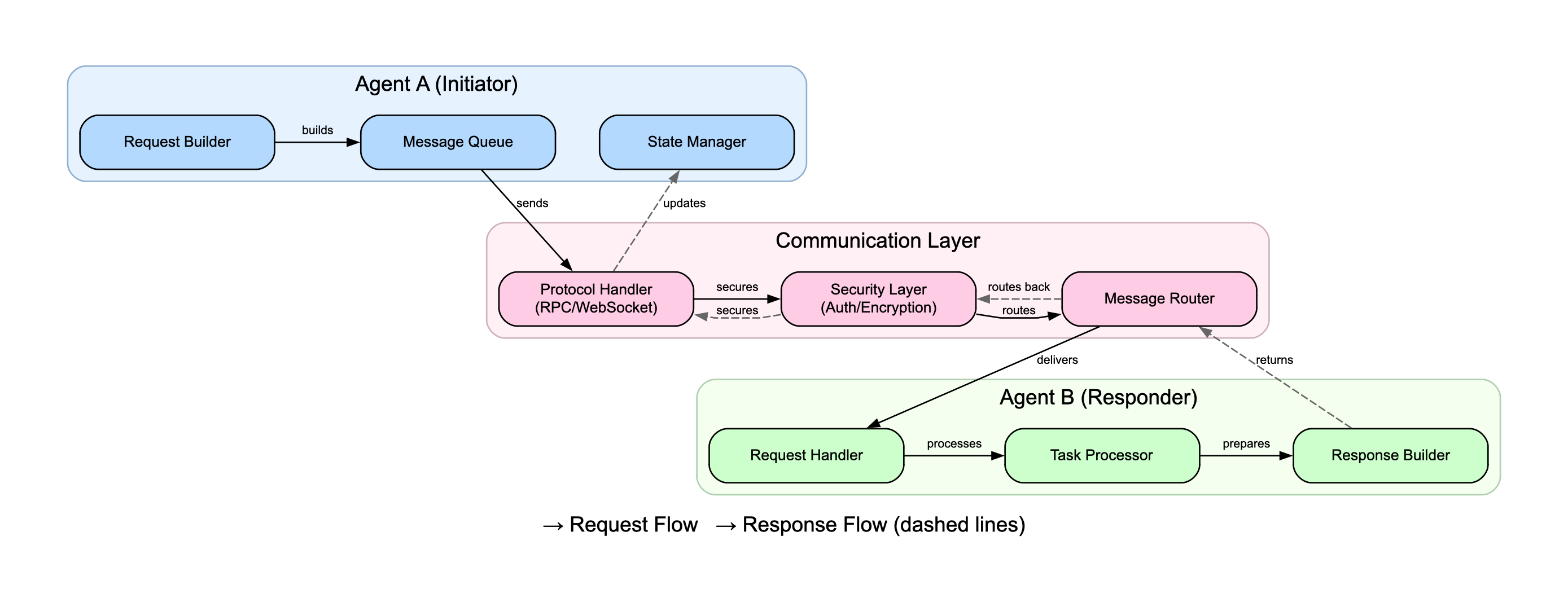}
    \vspace{-30pt}
    \caption{Agent-agent communication over the AIOS server communication layer.}
    \label{fig:agent_agent_interaction}
\end{figure}

\begin{figure}[t]
    \centering
    \includegraphics[width=0.9\linewidth]{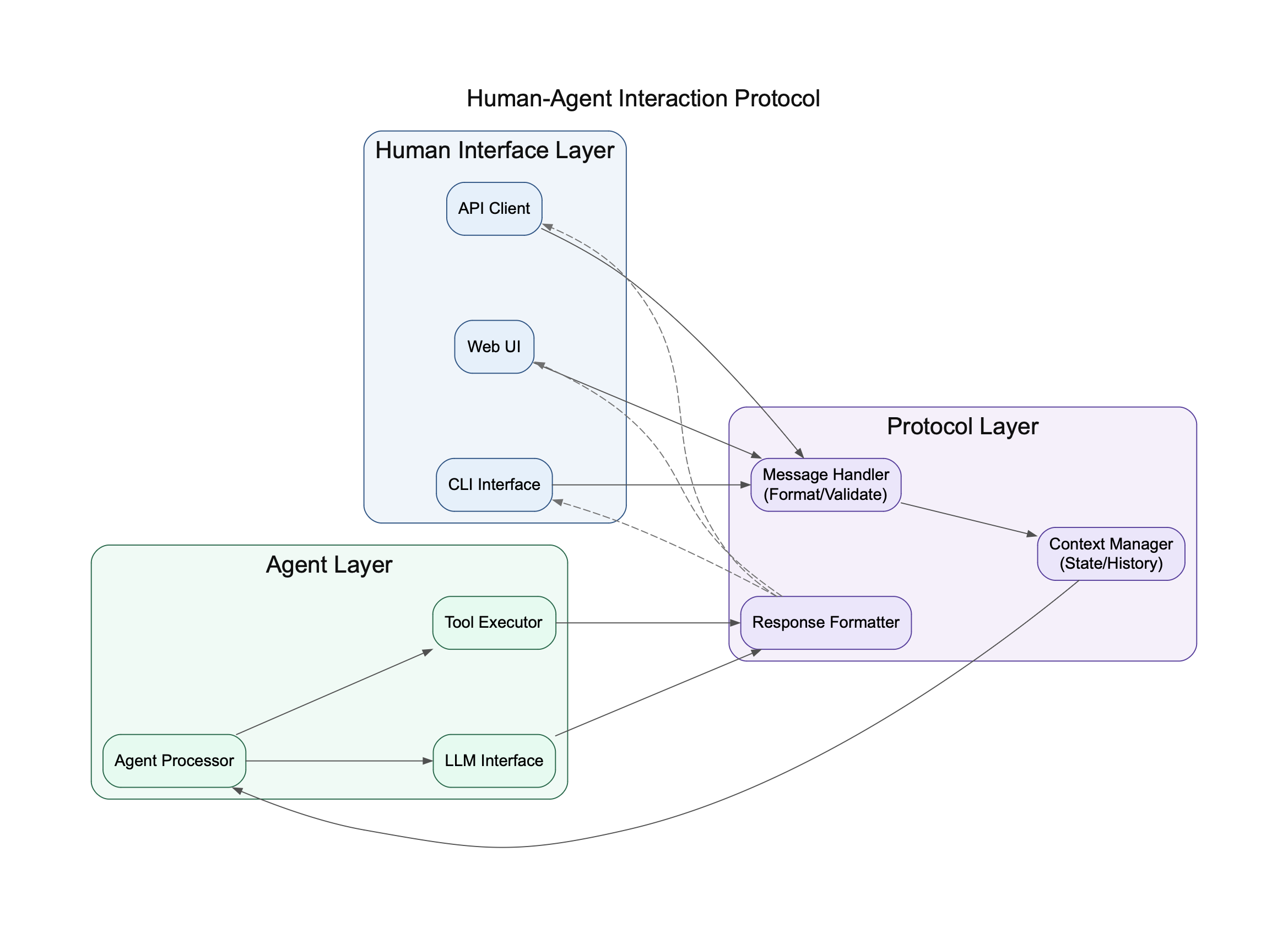}
    \vspace{-25pt}
    \caption{Human-agent communication via user interface.}
    \label{fig:human_agent_interaction}
\end{figure}

Figure~\ref{fig:agent_agent_interaction} and Figure~\ref{fig:human_agent_interaction} illustrate the primary communication modes in the AIOS server system: agent-agent communication and human-agent communication.

Figure~\ref{fig:agent_agent_interaction} depicts the agent-to-agent messaging workflow. The initiating agent constructs a request via the Request Builder and sends it through a protocol layer that supports RPC or WebSocket. The message is authenticated and encrypted before being routed to the target agent. The receiving agent processes the request and returns a structured response via the same secure channel.

In contrast, Figure~\ref{fig:human_agent_interaction} presents the human-agent interaction pipeline. Human users can interact through multiple interfaces (API, CLI, or Web UI). The protocol layer formats, validates and enriches the message with context before passing it to the agent layer, where internal tools or LLMs execute the task. Responses are formatted and routed back through the same protocol layer.

These interaction flows demonstrate the modularity and interoperability of the AIOS server communication protocols, supporting machine-to-machine coordination and user-facing automation.

\section{Node Registration and Discovery}


Scalable agent communication requires effective methods for agent discovery and management.
AIOS server addresses this challenge by implementing a decentralized agent registration and discovery mechanism, ensuring a safe, fault-tolerant, and adaptive agent ecosystem.

Besides nodes that host agents, there are one or more agent registry nodes in the network. When an agent node is launched, it needs to register itself on one or more registry nodes, so that it can be discovered and receive tasks in the network. After registration, each agent node periodically advertises its availability using a structured metadata which includes the following fields:
\begin{itemize}
    \item \texttt{"agent\_id"}: Unique identifier of the agent, typically in the format \texttt{namespace/agent\_name}.
    \item \texttt{"description"}: List of functional tags or capabilities offered by the agent.
    \item \texttt{"last\_seen"}: Timestamp (in UTC) indicating the most recent broadcast from the agent.
\end{itemize}

\begin{jsonlisting}[title={Example of Agent Metadata for Availability Broadcast}]
{
  "agent_id": "example/academic_agent",
  "description": ["text_analysis", "research"],
  "last_seen": "2025-02-28T12:00:00Z"
}
\end{jsonlisting}

This decentralized agent registration and discovery mechanism eliminates reliance on a single registration service and improves system resilience, meanwhile enhance the safety of the network since agent information can be found on one or more registry nodes to prevent malicious agents in the network. We provide the design and implementation details of the mechanism in the following.

\subsection{AIOS Server as an Autonomous Node}

AIOS Server functions as independent, self-regulating entities capable of dynamic task delegation and workload distribution. Each server node serves a dual role: it acts as both a \textit{service provider} by hosting agents and a \textit{dynamic client} by accessing external services.
Each node supports: \textit{Service Hosting}: AIOS server node exposes API endpoints for agent-based task execution, and \textit{Remote Invocation}: Nodes delegate tasks to other AIOS agents when needed. This structure allows for adaptive workload balancing and efficient inter-node communication.

When a node receives a task request, it follows the following \textit{Agent Execution Workflow}:
\begin{enumerate}[leftmargin=*]
    \item \textit{Local Execution}: The task is executed locally if a suitable agent is available.
    \item \textit{Task Delegation}: If no local agent is available, the task is delegated to another AIOS server node via an \textit{adaptive routing} mechanism.
    \item \textit{Task Completion}: The designated agent processes the request and returns the result.
    \item \textit{Result Integration}: The originating node receives and integrates the response.
\end{enumerate}


Each node periodically reports its state and active agents. Examples for the structure of AIOS node status report is shown in Appendix~\ref{appendix:aios-node-report}, and the structure of AIOS node task assignment is shown in Appendix~\ref{appendix:aios-task-assignment}.


The AIOS autonomous node architecture provides several benefits: 1) \textit{Scalability}: Nodes operate independently, enabling seamless system expansion; 2) \textit{Fault Tolerance}: Failure of a single node does not impact system functionality; 3) \textit{Load Balancing}: Tasks are dynamically allocated based on node capacity; 4) \textit{Decentralized Execution}: AIOS nodes reduce reliance on static configurations.

\subsection{Decentralized Registration with Distributed Hash Table and Gossip Protocol}

We design and implement a decentralized registration system to support scalable, fault-tolerant agent discovery in the ecosystem. This system integrates a Distributed Hash Table (DHT) and a Gossip-based synchronization protocol, enabling AIOS agents to register, discover, and monitor each other across a peer-to-peer (P2P) network without relying on centralized services.


The prototype architecture integrates two core components to support decentralized agent registration and synchronization:

\begin{itemize}
    \item \textit{Kademlia-based Distributed Hash Table (DHT)}: Provides structured, key-based metadata storage and lookup with logarithmic complexity $O(\log n)$ across $n$ nodes.
    \item \textit{Gossip-based Synchronization Protocol}: Enables periodic, peer-to-peer propagation of agent presence and status updates with eventual consistency guarantees.
\end{itemize}

These components are exposed to the AIOS server runtime through an abstraction layer, allowing seamless integration with higher-level agent workflows. This design separates protocol logic from task execution, enabling modular deployment and interoperability.

\begin{figure}[t]
    \centering
    \includegraphics[width=1\linewidth]{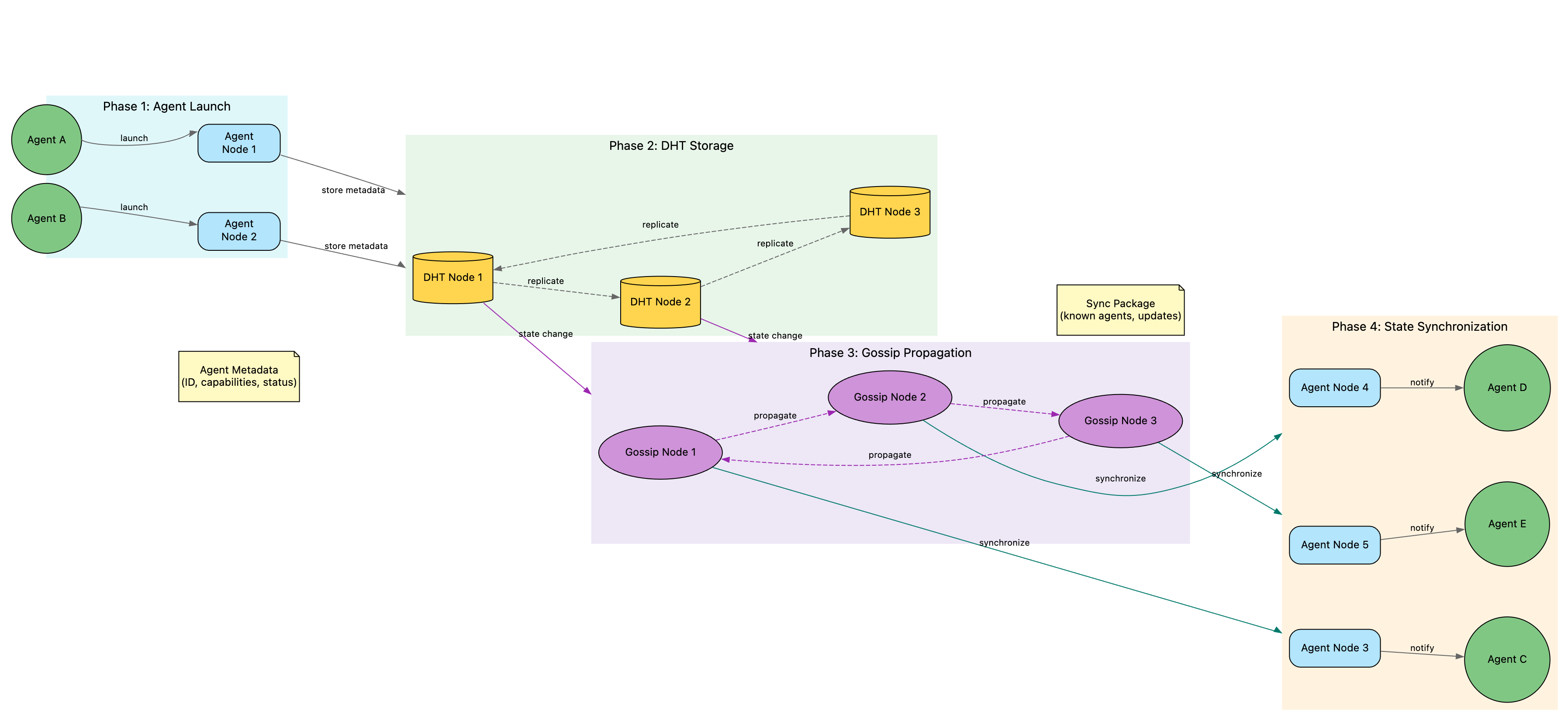}
    \caption{Decentralized agent discovery and metadata propagation pipeline. The system operates in four stages: (1) agents are launched with local agent nodes; (2) metadata is stored and replicated across neighboring nodes via DHT; (3) presence and state changes are propagated using the Gossip protocol; (4) other agent nodes synchronize state and notify agents of network updates.}
    \label{fig:dht-gossip-flow}
\end{figure}

Figure~\ref{fig:dht-gossip-flow} presents the end-to-end metadata flow in the decentralized system, highlighting the hybrid interplay between DHT-based storage and Gossip-driven state dissemination. Note that although there are different types of nodes in the system, all of the nodes are basically running the same AIOS server code base, they are just taking different roles by activating different functionalities in the sever. The architecture avoids centralized coordination while ensuring scalability, fault tolerance, and high metadata availability under dynamic network conditions.

The overall process unfolds through four stages:

\begin{enumerate}[leftmargin=*]
    \item \textit{Agent Launch}: Agents are launched on local AIOS server node and publish their capability metadata to the node.
    \item \textit{Agent Registration on DHT Storage}: The DHT nodes store and replicate metadata across the DHT overlay for fault-tolerant lookup.
    \item \textit{Gossip Dissemination}: Gossip nodes periodically exchange presence and status deltas with a random subset of peers.
    \item \textit{State Synchronization}: Other agent nodes apply received updates and notify their agents of topology or role changes.
\end{enumerate}

This decentralized design enables robust, scalable agent discovery and coordination. It remains resilient to node churn and transient failures, offering a practical foundation for distributed, multi-agent collaboration in real-world deployment environments. Detailed implementation logic and source code examples are provided in Appendix~\ref{appendix:dht-gossip-implementation}.

\subsection{Functional Capabilities and Design Trade-offs}

In summary, the decentralized registration system supports the following key functions: 1) \textit{Agent Registration and Lookup}: Agents publish their identity and capability metadata to the DHT using globally unique keys. Peers can efficiently retrieve these records to locate suitable collaborators; 2) \textit{Dynamic Node Management}: New nodes can join the system anytime, synchronize routing tables, and begin participating in the registration network with minimal configuration; 3) \textit{Fault-Tolerant Metadata Replication}: The system replicates agent records across multiple DHT nodes to maintain availability during node failures or temporary disconnections; 4) \textit{Real-Time Presence Dissemination}: Using the Gossip protocol, AIOS nodes periodically exchange state information to track agent liveness and capability updates in near real-time.


This design enables several important advantages of the system: 1) \textit{High Availability}: No single point of failure---nodes can join or leave without disrupting global service; 2) \textit{Scalability}: Performance scales logarithmically with network size due to the DHT structure; 3) \textit{Self-Organization}: Nodes form an adaptive topology, requiring no centralized coordination; 4) \textit{Resilient Status Tracking}: Gossip-based synchronization ensures soft-state convergence even under unreliable connectivity.




\subsection{Distributed Agent Hub}

\begin{figure}[t]
    \centering
    \includegraphics[width=0.9\linewidth]{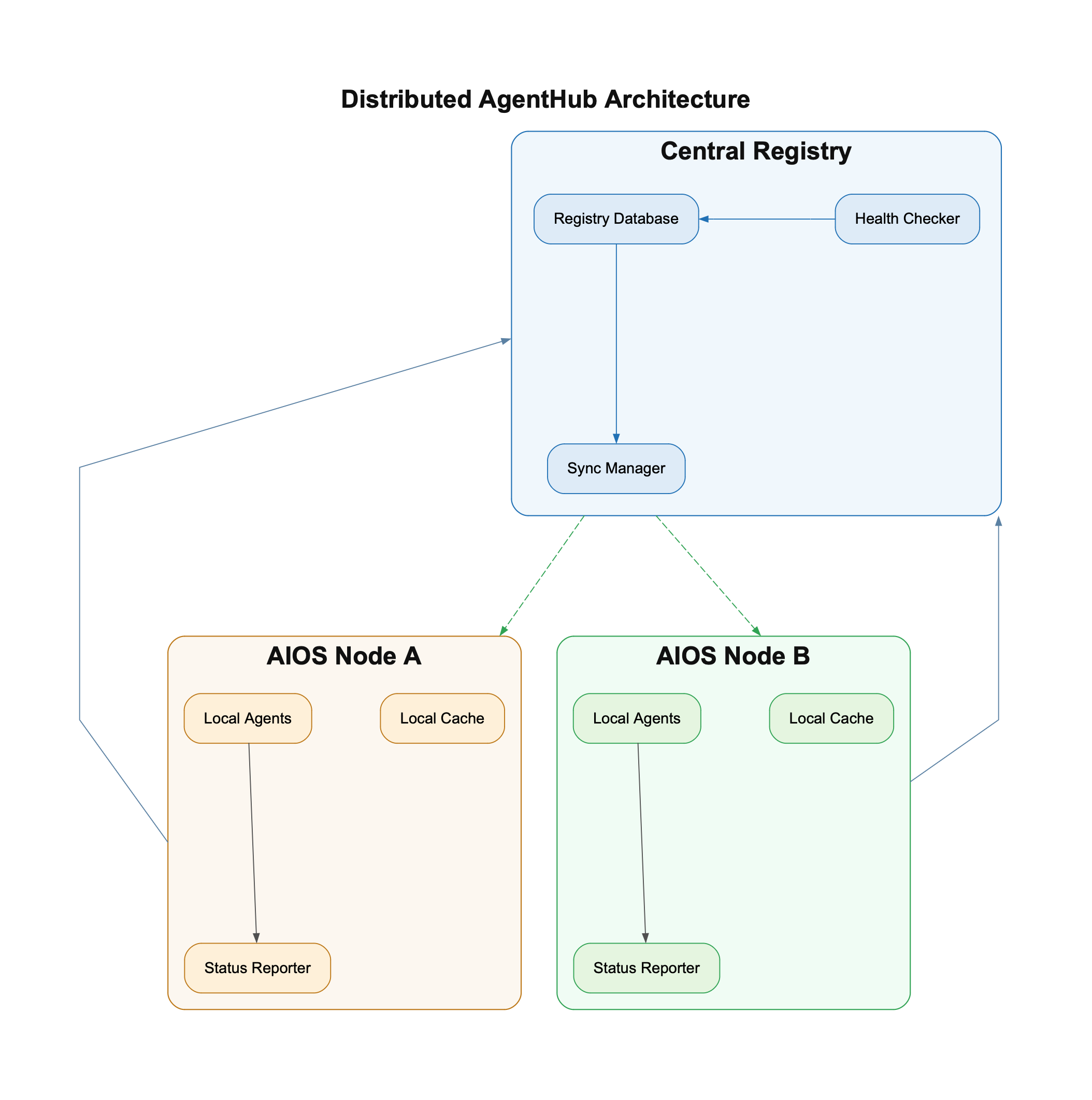}
    \vspace{-20pt}
    \caption{Distributed Agent Hub for decentralized agent management.}
    \label{fig:distributed_agenthub}
    \vspace{-10pt}
\end{figure}

Figure~\ref{fig:distributed_agenthub} illustrates the architecture of the distributed AgentHub system. Each AIOS agent node maintains a set of local agents and a local cache, reporting its status to the central registry nodes. The central registry nodes manage a global view of agent availability through a registry database and monitors node health using a dedicated checker. A synchronization manager propagates updates across the network, ensuring consistency between nodes. This design supports decentralized agent discovery, fault-tolerant registration, and real-time coordination across geographically distributed nodes. Table \ref{tab:aios_architecture_comparison} shows a comparison between the centralized client-server mode and the decentralized internet of agentsite mode of AIOS.

\begin{table}[t]
\centering
\begin{tabular}{|c|c|c|}
\hline
\textbf{Feature} & \textbf{Centralized AIOS} & \textbf{Decentralized AIOS} \\
\hline
\textbf{Agent Registration} & Managed by a central server & Distributed across AIOS nodes \\
\hline
\textbf{Fault Tolerance} & Single point of failure & Resilient via peer-to-peer network \\
\hline
\textbf{Scalability} & Limited by central server capacity & Horizontally scalable \\
\hline
\textbf{Task Delegation} & Static assignment & Dynamic inter-node routing \\
\hline
\textbf{Message Routing} & Centralized relay & Multi-hop decentralized routing \\
\hline
\textbf{Example Use Case} & Small-scale AI automation & Large-scale distributed AI collaboration \\
\hline
\end{tabular}
\caption{Comparison of Centralized and Decentralized AIOS Architectures}
\label{tab:aios_architecture_comparison}
\end{table}

\section{Experiments}

We evaluate the AIOS server communication framework under local and cloud deployments. Experiments assess three core metrics: latency, throughput, and agent registration efficiency in a decentralized setting.

\subsection{Experimental Setup}

All tests were performed using structured JSON-RPC requests under controlled concurrency levels. Two environments were evaluated:

\begin{itemize}
    \item \textbf{Local}: Simulated on a macOS machine (Apple M-series CPU) running multiple AIOS nodes.
    \item \textbf{Cloud}: Deployed to a public endpoint at \url{https://planet.aios.foundation}.
\end{itemize}

Each test involved 50, 100, and 200 total requests issued through 5, 10, and 20 concurrent users, respectively. 

\subsection{Communication Performance Results}

AIOS server achieved 100\% response success in all scenarios. Latency scaled predictably with load, while throughput increased steadily. The system demonstrated reliable performance in both local and cloud environments. Despite infrastructure differences, latency remained under 200ms in all cases. Throughput increased with load, reaching up to 229 requests per second in the cloud deployment. This confirms the framework's scalability and consistency across platforms.

\begin{table}[t]
\centering
\begin{tabular}{|c|c|c|c|}
\hline
\textbf{Environment} & \textbf{Load} & \textbf{Avg. Latency (s)} & \textbf{Throughput (req/s)} \\
\hline
Local & 50 reqs & 0.061 & 80.2 \\
Local & 100 reqs & 0.104 & 116.9 \\
Local & 200 reqs & 0.165 & 163.1 \\
\hline
Cloud & 50 reqs & 0.143 & 32.2 \\
Cloud & 100 reqs & 0.145 & 100.0 \\
Cloud & 200 reqs & 0.145 & 229.3 \\
\hline
\end{tabular}
\caption{AIOS server communication performance under local and cloud deployments}
\label{tab:comm_perf}
\end{table}

\subsection{Node Discovery Performance}

The agent discovery protocol was tested using distributed AIOS nodes and evaluated based on registration latency. Across 3, 5, and 7 nodes, all agents successfully registered within milliseconds. Average registration time remained consistent at 1ms, with a maximum delay of 2ms. These results confirm the efficiency and stability of the DHT and gossip-based synchronization mechanism.
Overall, the AIOS server protocol demonstrates low-latency, high-throughput communication and rapid agent registration across diverse environments. These findings confirm its robustness for decentralized agent communication.

\subsection{System Visualization and Demonstration}

This section presents interface-level illustrations of the AIOS server infrastructure. These visuals support the system's functional claims, highlighting its distributed design, agent orchestration, and human-agent interaction.

\subsubsection{Global Node Distribution}

\begin{figure}[H]
  \centering
  \begin{subfigure}[b]{0.49\textwidth}
    \centering
    \includegraphics[width=\textwidth, height=0.32\textheight]{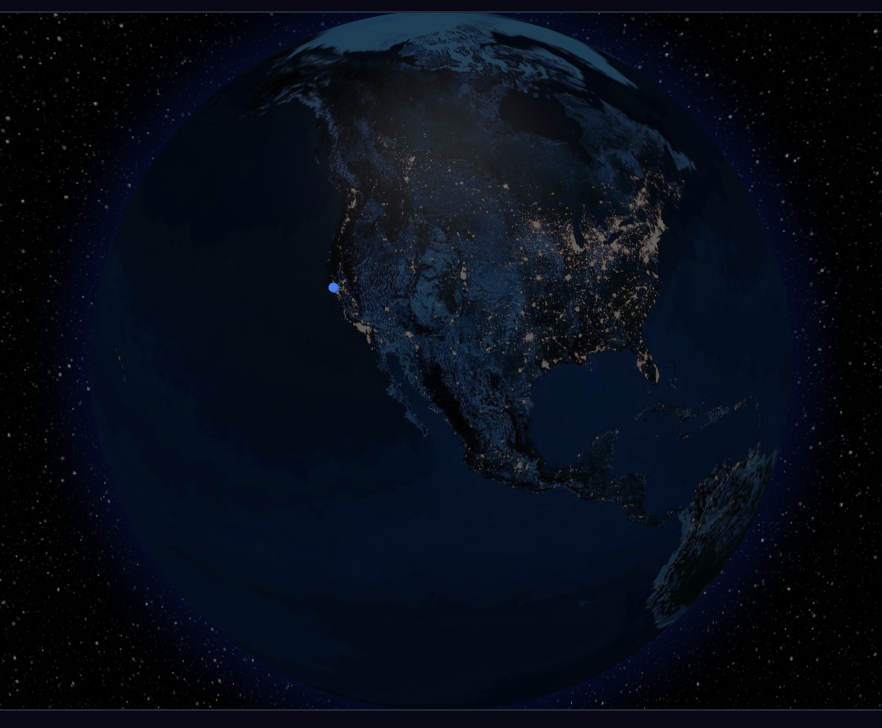}
    \caption{Pacific view showing active nodes in San Francisco.}
    \label{fig:node_map_pacific}
  \end{subfigure}
  \hfill
  \begin{subfigure}[b]{0.49\textwidth}
    \centering
    \includegraphics[width=\textwidth, height= 0.32\textheight]{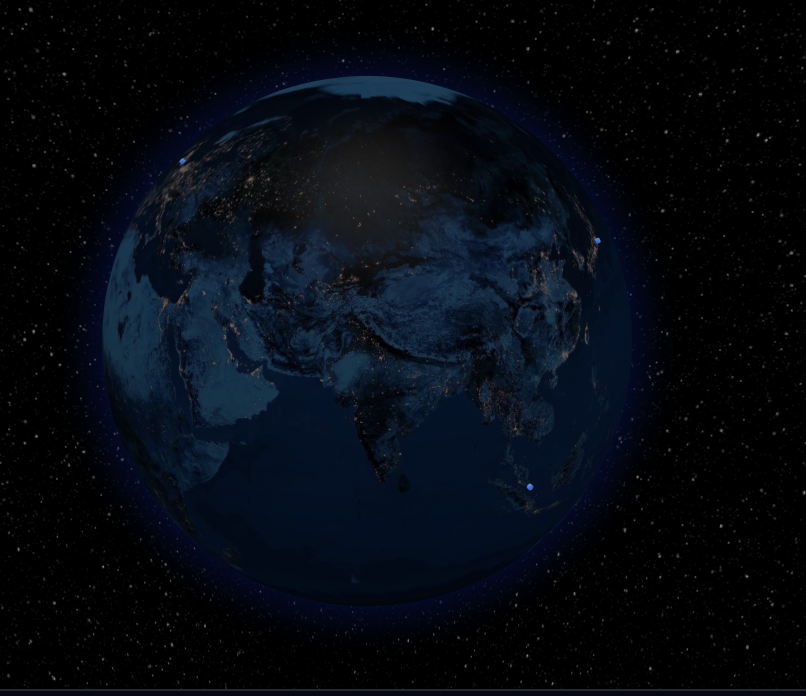}
    \caption{Eurasian view showing active nodes in London, Singapore, Tokyo.}
    \label{fig:node_map_eurasia}
  \end{subfigure}
  \caption{Global nodes map showing nodes at different locations across the Internet of Agentsites.}
\end{figure}

Figures~\ref{fig:node_map_pacific} and~\ref{fig:node_map_eurasia} display the global distribution of active AIOS nodes. Each point represents an autonomous server instance hosting LLM agents. These nodes are geographically dispersed and form a decentralized agent network.



\subsubsection{Node Overview Interface}

\begin{figure}[H]
    \centering
    \includegraphics[width=1.0\linewidth]{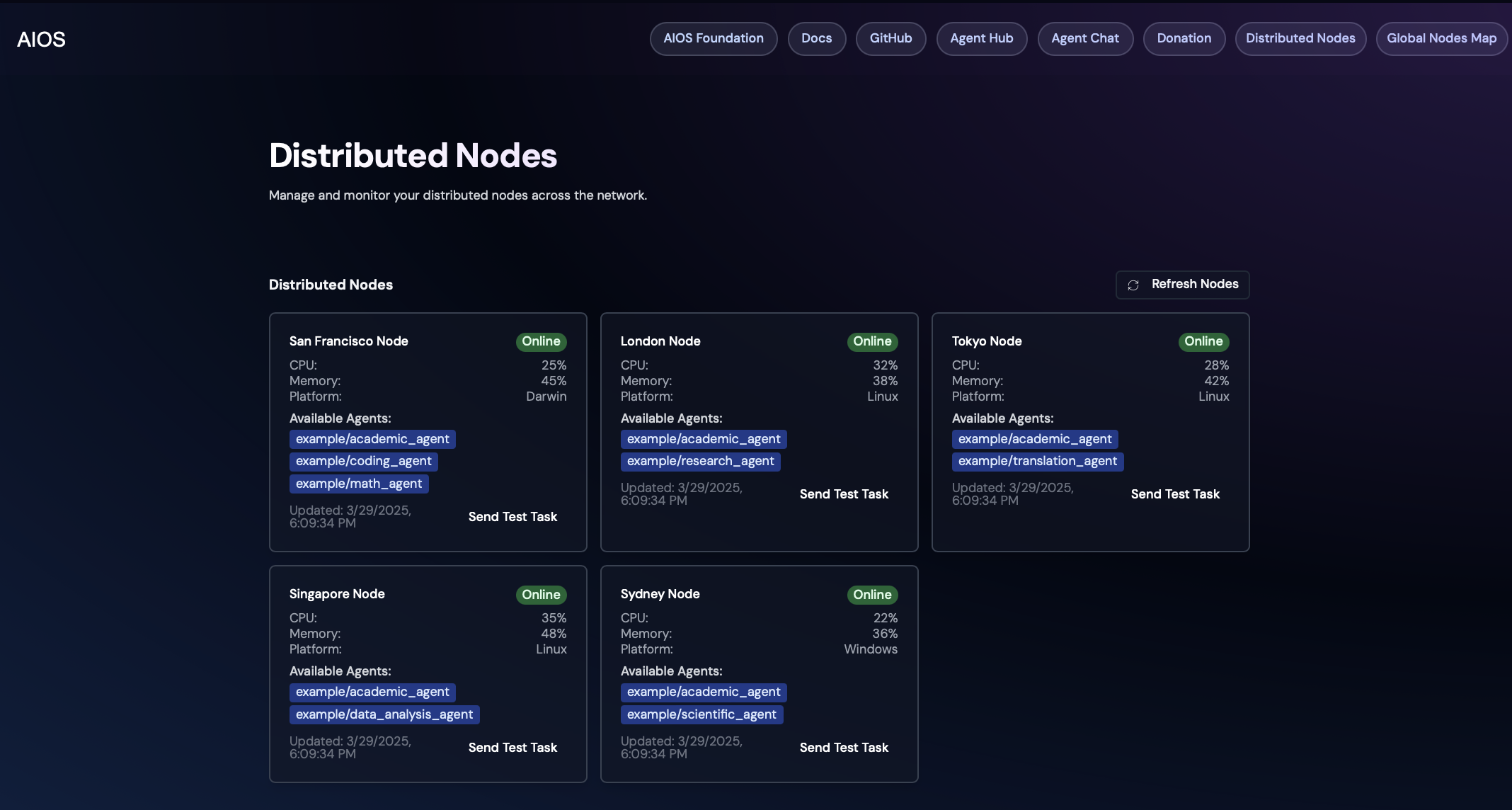}
    \caption{Distributed node dashboard showing CPU/memory usage, platform type, agent availability.}
    \label{fig:node_list_dashboard}
\end{figure}

Figure~\ref{fig:node_list_dashboard} provides a snapshot of the AIOS node dashboard. Each node card shows its current resource usage, platform type, and available agents. All nodes are synchronized and support real-time task execution.

\subsubsection{Node Detail and Interaction Interface}

\begin{figure}[H]
    \centering
    \includegraphics[width=1.0\linewidth]{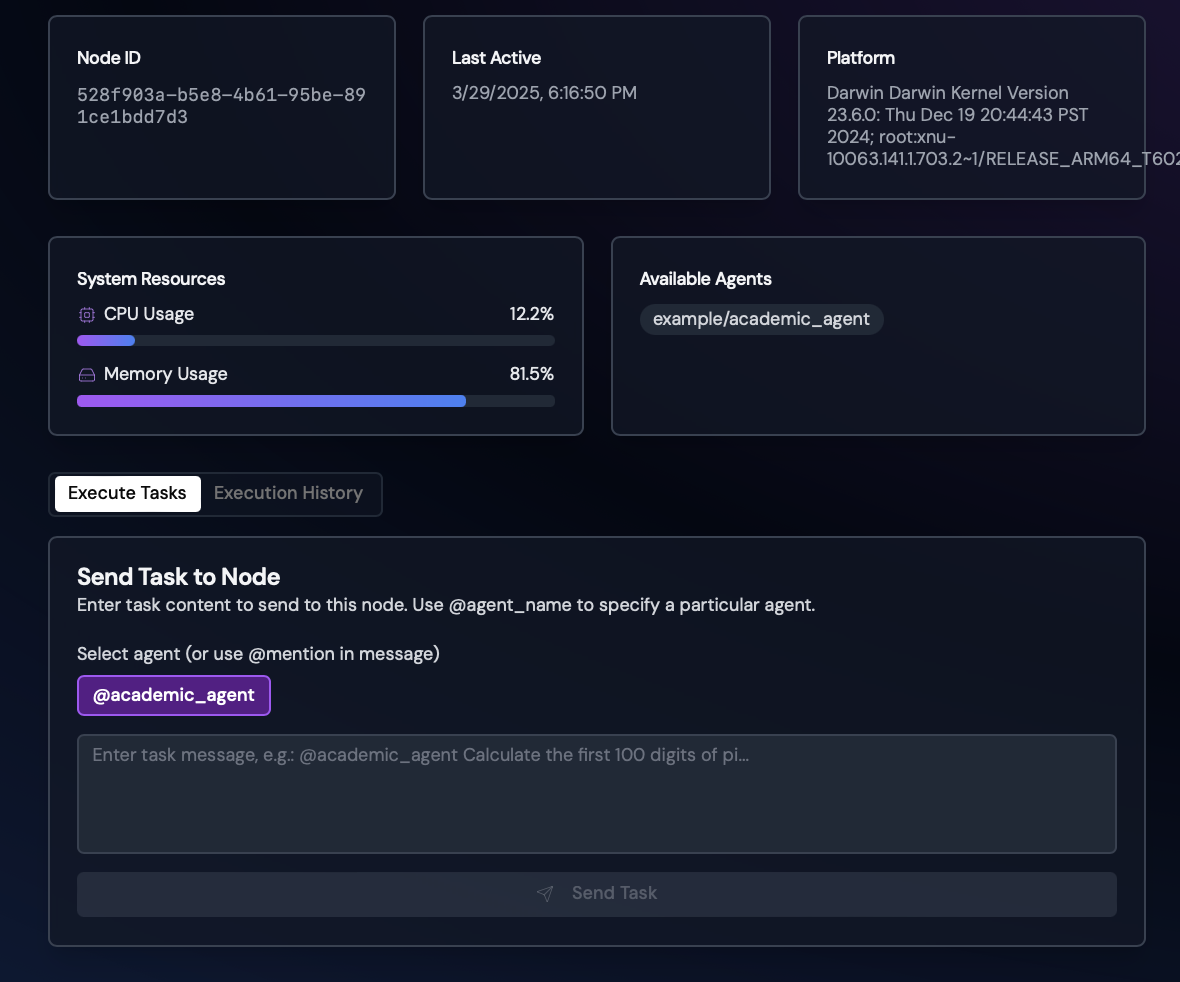}
    \caption{Interface of the San Francisco node with task input, resource usage, and agent selector.}
    \label{fig:single_node_view}
\end{figure}

Figure~\ref{fig:single_node_view} presents the detailed view of an individual AIOS node. Users can inspect system performance, choose specific agents, and issue tasks to the agents hosted on the node via the UI.

\subsubsection{Task Execution and Result Logging}

\begin{figure}[t]
    \centering
    \includegraphics[width=1.0\linewidth]{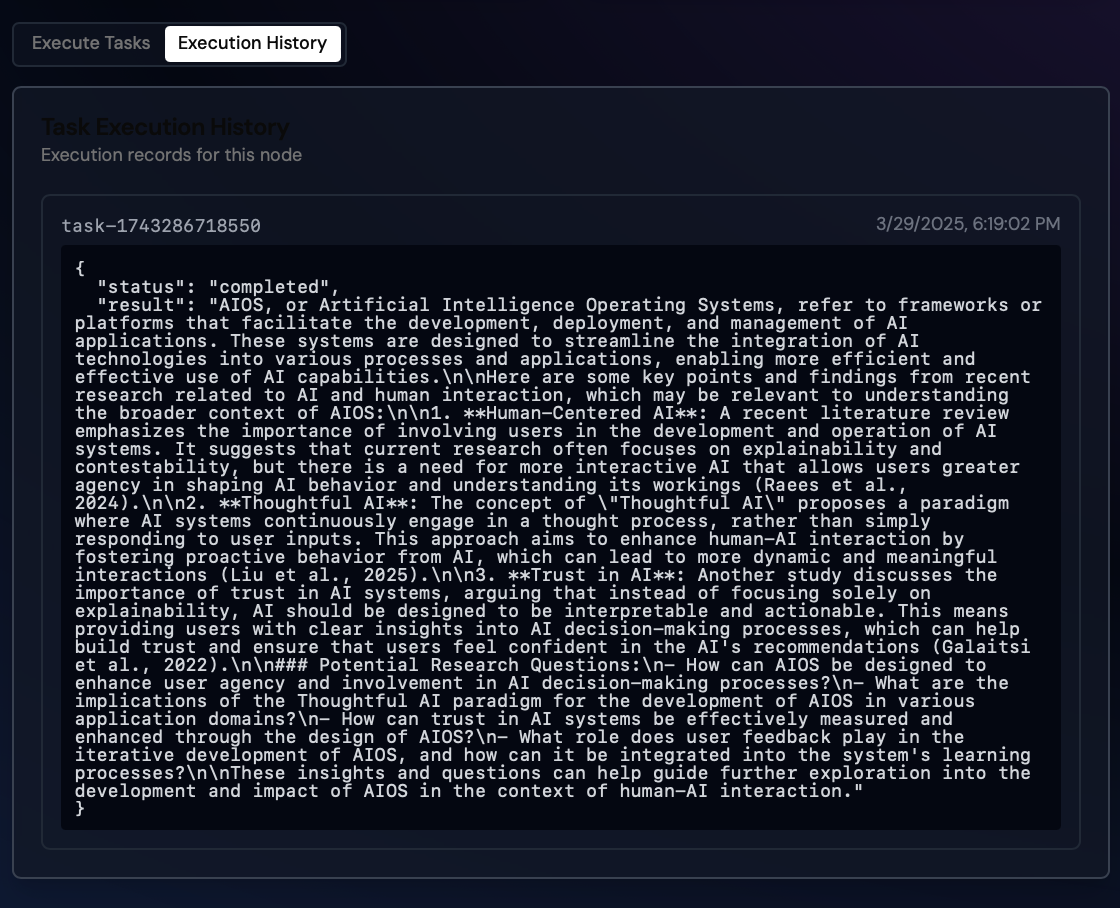}
    \caption{Agent response log showing a completed task with structured output.}
    \label{fig:task_execution_result}
\end{figure}

Figure~\ref{fig:task_execution_result} shows a completed task handled by the \texttt{academic\_agent}. The interface logs the complete JSON-RPC request and the corresponding AI-generated response. This validates the agent's autonomous reasoning capability and system traceability.

\section{Conclusions and Future Work}

This work presents AIOS server, a foundational infrastructure for the internet of agentsites, where each server node can host agents to enable task solving at global scale. One key advantage of AIOS server lies in its modular, message-oriented architecture, which supports flexible agent orchestration across heterogeneous environments. The structured communication protocol (MCP + JSON-RPC) further facilitates interoperability between agents and external services. Evaluation demonstrates that the AIOS server protocol achieves low latency, high throughput, and efficient decentralized agent registration under local and cloud-based deployments. These results validate its viability for real-world applications, particularly in single-agent or loosely coupled multi-agent systems.

Future development will focus on extending AIOS to support multi-agent task orchestration and improving its robustness in more complex, failure-prone settings, including: 1) \textit{Inter-node communication optimization}: Reducing serialization overhead and supporting asynchronous messaging to improve throughput under high concurrency; 2) \textit{Adaptive load balancing}: Developing real-time node profiling and routing strategies based on resource availability and agent capabilities; 3) \textit{Resilience under partial failure}: Introducing fallback mechanisms, timeout-based task migration, and state checkpointing to improve fault tolerance; 4) \textit{DHT registry validation at scale}: Integrating the DHT + Gossip registry into a larger distributed deployment with hundreds of nodes and evaluating consistency and convergence under dynamic network conditions; 5) \textit{Security and trust modeling}: Designing lightweight authentication and authorization schemes for peer-to-peer agent communication.


\section*{Acknowledgment}
The \LaTeX~template used in this paper is adapted from the open-source template developed by Rongsheng Wang. We thank him for his contributions and efforts.

\bibliography{IoA}

\newpage

\appendix

\section*{APPENDIX}

\section{Human-Agent Communication}
\label{appendix:human-agent}

\subsection{Message Schema Explanation}
The following explains each field of the human-agent communication request:

\begin{itemize}
    \item \texttt{"jsonrpc"}: Protocol version identifier (e.g., "2.0").
    \item \texttt{"id"}: Unique identifier used to match requests and responses.
    \item \texttt{"method"}: Name of the remote method invoked (e.g., \texttt{"aios/delegateTask"}).
    \item \texttt{"params"}: Contains task-specific parameters:
    \begin{itemize}
        \item \texttt{"sender"}: Entity initiating the request.
        \item \texttt{"recipient"}: Target AI agent receiving the request.
        \item \texttt{"messages"}: User-provided messages forming the input.
        \item \texttt{"maxTokens"}: Token limit for the response output.
    \end{itemize}
\end{itemize}

\subsection{Example: User Request}
\label{appendix:human-agent-request}

\begin{jsonlisting}[title={Example: User Request}]
{
    "jsonrpc": "2.0",
    "id": "12345",
    "method": "aios/delegateTask",
    "params": {
        "sender": {
            "id": "human_user"
        },
        "recipient": {
            "id": "academic_agent",
            "role": "assistant"
        },
        "messages": [
            {
                "role": "user",
                "content": {
                    "type": "text",
                    "text": "Summarize the key findings of this research paper."
                }
            }
        ],
        "maxTokens": 200
    }
}
\end{jsonlisting}

\subsection{Message Schema Explanation (Response)}
\begin{itemize}
    \item \texttt{"jsonrpc"}: Protocol version.
    \item \texttt{"id"}: Corresponds to the original request ID.
    \item \texttt{"result"}: Contains the response from the agent:
    \begin{itemize}
        \item \texttt{"sender"}: Responding agent.
        \item \texttt{"recipient"}: Original human user.
        \item \texttt{"content"}: Task result or response content.
        \item \texttt{"model"}: Language model used to generate response.
        \item \texttt{"stopReason"}: Indicates how response was terminated.
    \end{itemize}
\end{itemize}

\subsection{Example: AI Agent Response}
\label{appendix:human-agent-response}

\begin{jsonlisting}[title={Example: AI Agent Response}]
{
    "jsonrpc": "2.0",
    "id": "12345",
    "result": {
        "sender": {
            "id": "academic_agent",
            "role": "assistant"
        },
        "recipient": {
            "id": "human_user"
        },
        "content": {
            "type": "text",
            "text": "The paper presents a novel approach to optimizing transformer models, achieving a 20% reduction in computational cost while maintaining accuracy."
        },
        "model": "GPT-4",
        "stopReason": "endTurn"
    }
}
\end{jsonlisting}

\newpage

\section{Agent-Agent Communication}
\label{appendix:agent-agent}

\subsection{Message Schema Explanation (Request)}
\begin{itemize}
    \item \texttt{"jsonrpc"}: Protocol version.
    \item \texttt{"id"}: Identifier of the task.
    \item \texttt{"method"}: Operation being requested (e.g., \texttt{"aios/delegateTask"}).
    \item \texttt{"params"}:
    \begin{itemize}
        \item \texttt{"intent"}: Purpose of the request (e.g., data extraction).
        \item \texttt{"sender"}: Requesting agent and role.
        \item \texttt{"recipient"}: Target agent and role.
        \item \texttt{"task"}: Task metadata and parameters.
    \end{itemize}
\end{itemize}

\subsection{Task Delegation Request}
\label{appendix:agent-request}

\begin{jsonlisting}[title={Task Delegation Request}]
{
    "jsonrpc": "2.0",
    "id": "task-001",
    "method": "aios/delegateTask",
    "params": {
        "intent": "extract_data",
        "sender": {
            "id": "analysis_agent",
            "role": "processor"
        },
        "recipient": {
            "id": "data_agent",
            "role": "retriever"
        },
        "task": {
            "name": "Extract statistical features",
            "arguments": {
                "dataset": "financial_reports_2023.csv",
                "features": ["mean", "std", "sample_size"]
            }
        }
    }
}
\end{jsonlisting}

\subsection{Message Schema Explanation (Response)}
\begin{itemize}
    \item \texttt{"jsonrpc"}: Protocol version.
    \item \texttt{"id"}: Same ID as the original request.
    \item \texttt{"result"}:
    \begin{itemize}
        \item \texttt{"sender"}: Agent responding to the task.
        \item \texttt{"recipient"}: Agent that issued the task.
        \item \texttt{"content"}: Result of the task, including output.
        \item \texttt{"isError"}: Indicates if the task failed.
    \end{itemize}
\end{itemize}

\subsection{Agent Response}
\label{appendix:agent-response}

\begin{jsonlisting}[title={Agent Response}]
{
    "jsonrpc": "2.0",
    "id": "task-001",
    "result": {
        "sender": {
            "id": "data_agent",
            "role": "retriever"
        },
        "recipient": {
            "id": "analysis_agent",
            "role": "processor"
        },
        "content": {
            "task": "Extract statistical features",
            "status": "completed",
            "output": {
                "mean": 85.3,
                "std": 4.2,
                "sample_size": 500
            }
        },
        "isError": false
    }
}
\end{jsonlisting}

\newpage
\section{AIOS Node Communication Formats}
\label{appendix:aios-communication}

\subsection{AIOS Node Status Report}
\label{appendix:aios-node-report}

\paragraph{Field Explanation:}
\begin{itemize}
    \item \texttt{"node\_id"}: Unique identifier for the node.
    \item \texttt{"node\_name"}: System name of the node.
    \item \texttt{"timestamp"}: UTC timestamp of the report.
    \item \texttt{"system\_info"}: Hardware and usage statistics.
    \item \texttt{"available\_agents"}: List of agents currently deployed on the node.
\end{itemize}

\begin{jsonlisting}[title={AIOS Node Status Report}]
{
  "node_id": "Node_42",
  "node_name": "aios-compute-1",
  "timestamp": "2025-02-28T12:00:00Z",
  "system_info": {
    "cpu_percent": 23.4,
    "memory_percent": 67.2,
    "platform": "Linux"
  },
  "available_agents": [
    "example/academic_agent",
    "example/math_agent",
    "example/language_tutor"
  ]
}
\end{jsonlisting}

\subsection{AIOS Task Assignment Format}
\label{appendix:aios-task-assignment}

\paragraph{Field Explanation:}
\begin{itemize}
    \item \texttt{"task\_id"}: Task identifier.
    \item \texttt{"assigned\_agent"}: Name of the agent responsible for the task.
    \item \texttt{"status"}: Current execution status (e.g., running, completed).
\end{itemize}

\begin{jsonlisting}[title={AIOS Node Task Assignment}]
{
  "task_id": "task-987",
  "assigned_agent": "example/language_tutor",
  "status": "running"
}
\end{jsonlisting}

\newpage

\section{Implementation of Distributed Agent Registry Prototype}
\label{appendix:dht-gossip-implementation}

This section presents key excerpts from the prototype implementation of AIOS's decentralized agent registration and discovery system. The system combines a Kademlia-based Distributed Hash Table (DHT) for structured key-value storage with a lightweight Gossip protocol for periodic presence synchronization.

\subsection{DHT Class for Agent Registration}

This class handles agent metadata registration into the DHT overlay, enabling global discoverability via unique identifiers. Metadata such as node address and timestamp is encoded before being distributed.

\begin{lstlisting}[language=Python, caption={DHT Agent Registry Class}]
class DHT:
    def __init__(self, ip, port, node_id=None, k=20):
        self.node_id = node_id or NodeID()
        self.ip = ip
        self.port = port
        self.local_node = Node(self.node_id, ip, port)
        self.routing_table = RoutingTable(self.node_id, k)
        self.data_store = {}
        
    def register_agent(self, agent_id, metadata):
        key = f"agent:{agent_id}"
        metadata["last_update"] = time.time()
        metadata["node_id"] = str(self.node_id)
        metadata["node_ip"] = self.ip
        metadata["node_port"] = self.port
        return self.store(key, metadata)
            
    def find_agent(self, agent_id):
        key = f"agent:{agent_id}"
        return self.lookup(key)
\end{lstlisting}

\subsection{Gossip Protocol for Presence Synchronization}

This module implements a gossip-based presence tracking protocol. Each node periodically propagates its knowledge of peer agents to a sampled subset of neighbors, balancing coverage and overhead.

\begin{lstlisting}[language=Python, caption={Presence Gossip Protocol}]
class GossipProtocol(asyncio.DatagramProtocol):
    def __init__(self, node_id, port=8001):
        self.node_id = node_id
        self.port = port
        self.peers = {}
        self.message_cache = {}
        self.callbacks = {}
        
    def _propagate_message(self, message):
        if message.ttl <= 1:
            return
            
        new_message = GossipMessage(
            message.sender_id,
            message.message_type,
            message.data,
            message.timestamp,
            message.ttl - 1
        )
        
        live_peers = [p for p in self.peers.values() 
                     if p["state"] != NodeState.DEAD]
                     
        if not live_peers:
            return
            
        target_count = min(len(live_peers), max(3, int(len(live_peers) ** 0.5)))
        targets = random.sample(live_peers, target_count)
        
        for peer in targets:
            addr = (peer["ip"], peer["port"])
            self._send_message(new_message, addr)
\end{lstlisting}

\subsection{Agent Directory Service}

This service provides high-level API access for agent capability-based queries. It relies on the underlying gossip protocol to maintain updated lists of agents and their advertised features.

\begin{lstlisting}[language=Python, caption={Agent Presence Directory}]
class GossipAgentDirectoryService:
    def __init__(self, node_id=None, host="127.0.0.1", port=8001, 
                seed_nodes=None):
        self.node_id = node_id or str(uuid.uuid4())
        self.presence_service = AgentPresenceService(self.node_id, host, port)
        self.seed_nodes = seed_nodes or []
        
    async def start(self):
        await self.presence_service.start()
        for node_id, host, port in self.seed_nodes:
            self.presence_service.add_peer(node_id, host, port)
        
    def register_agent(self, agent_id, capabilities=None):
        return self.presence_service.register_agent(agent_id, capabilities)
    
    def find_agents_by_capability(self, capability):
        agents = self.presence_service.get_agents_by_capability(capability)
        return [a.to_dict() for a in agents]
\end{lstlisting}

\subsection{Gossip Integrator for AIOS System}

This wrapper class integrates the gossip-based directory service into the broader AIOS runtime. It ensures that the decentralized components are initialized correctly and available via a consistent interface.

\begin{lstlisting}[language=Python, caption={Gossip Service Integrator}]
class GossipIntegrator:
    def __init__(self, config=None):
        self.config = config or Config()
        self.service = None
        self.node_id = self.config.get("p2p.node_id", default=None)
        self.host = self.config.get("p2p.gossip.host", default="127.0.0.1")
        self.port = self.config.get("p2p.gossip.port", default=8001)
        
    async def initialize(self):
        self.service = GossipAgentDirectoryService(
            node_id=self.node_id,
            host=self.host,
            port=self.port,
            seed_nodes=self.seed_nodes
        )
        await self.service.start()
        
    def register_agent(self, agent_id, capabilities=None):
        if not self.service:
            return False
        return self.service.register_agent(agent_id, capabilities)
\end{lstlisting}

\end{document}